\documentstyle[pre,multicol,aps,epsfig]{revtex}

\begin{document}

\title{Deciding the Nature of the ``Coarse Equation'' through
Microscopic Simulations: the Baby-Bathwater Scheme}

\author{Ju Li$^1$, Panayotis G. Kevrekidis$^2$, C. William Gear$^{3,4}$
and Ioannis G. Kevrekidis$^4$}

\address{$^1$Department of Materials Science and Engineering, Ohio
State University, Columbus, OH 43210}

\address{$^2$Department of Mathematics and Statistics, University
of Massachusetts, Amherst, MA 01003-4515}

\address{$^3$NEC Research Institute, 4 Independence Way,
Princeton, New Jersey 08540}

\address{$^4$Department of Chemical Engineering and PACM, Princeton
University, Princeton, New Jersey 08544}

\date{\today}
\maketitle

\begin{abstract}

Recent developments in multiscale computation allow the solution of
``coarse equations'' for the expected macroscopic behavior of
microscopically/stochastically evolving particle distributions without
ever obtaining these coarse equations in closed form.
The closure is obtained ``on demand'' through appropriately
initialized bursts of microscopic simulation.
The effective coupling of microscopic simulators with macrosocopic
behavior embodied in this approach requires certain decisions about
the nature of the unavailable ``coarse equation''.
Such decisions include (a) the determination of the highest spatial
derivative active in the equation, (b) whether the coarse equation
satisfies certain conservation laws, and (c) whether the coarse
dynamics is Hamiltonian.
These decisions affect the number and type of boundary conditions as
well as the nature of the algorithms employed in good solution
practice.
In the absence of an explicit formula for the temporal derivative, we
propose, implement and validate a simple scheme for deciding these and
other similar questions about the coarse equation using only the
microscopic simulator.
Microscopic simulations under periodic boundary conditions are carried
out for appropriately chosen families of random initial conditions;
evaluating the sample variance of certain statistics over the
simulation ensemble allows us to infer the highest order of spatial
derivatives active in the coarse equation.
In the same spirit we show how to determine whether a certain coarse
conservation law exists or not, and we discuss plausibility tests for
the existence of a coarse Hamiltonian or integrability.
We argue that such schemes constitute an important part of the equation-free
approach to multiscale computation.

\end{abstract}

\section{Introduction}
\label{sec:Introduction}

It is often the case that a microscopic or fine description of a
physical system is available, while we are interested in its
macroscopic or coarse behavior.
Consider, as an example, a biased random walk model for which the
particle density asymptotically evolves according to a macroscopic law
such as the Burgers equation.
Typically, the study of macroscopic behavior starts with obtaining a
closed PDE-level description (a ``coarse equation'') for the time
evolution of the expected or ensemble-averaged fields of a few, low
order moments of the micro-state phase space distribution.
For our example, this would be the zero-th moment, the density field. 
Then, an array of mathematical and computational tools (numerical
integration, fixed-point algorithms etc.) can be brought to bear on
the coarse equation.

Over the last few years, we have been developing a class of numerical
algorithms which attempt to analyze the coarse behavior without ever
obtaining the coarse equation in closed form \cite{Kevrekidis02}.
The common character of these schemes is to use short, appropriately
initialized bursts of microscopic simulations to estimate the
quantities which, if the coarse equation was available, we would
simply evaluate using the equation itself.
Such quantities, estimated on-demand, include the time derivative of
the evolving coarse fields, to be used in coarse projective
integration \cite{Gear01a}, or the effect of the time-evolution
operator for the implicit coarse Jacobian, to be used in Newton-Krylov
type contraction mappings like the Recursive Projection Method
(RPM) \cite{Shroff93,Gear02a} or in eigenvalue/vector computations.
These methods are based on matrix-free large scale scientific
computing, and we sometimes collectively refer to them as
``equation-free methods''.
What makes these computations possible is the assumption of a
separation of time scales in the dynamics of the evolving microscopic
distribution.
Typically, one finds that the hierarchy of coupled equations involving
higher cumulants of the microscopic distribution constitutes a
singularly perturbed problem: higher-order cumulants become, in the
course of simulation, quickly slaved to (become deterministic
functionals of) the lower-order cumulants.
The consequences of slaving, {\em realized in the computer as a
black-box}, embody the closures that allow us to solve for the coarse
behavior.
Fundamentally it is no different than if the closures are expressed in
closed form {\em first} and then evaluated {\em later}.
This way of thinking and newly developed computing technology can in
practice exceed the traditional approach in both accuracy and total
cost, especially if the constitutive relation is nonlinear and
multi-dimensional.
An example of this type is given in \cite{Li02}.
An additional advantage of such methods is their ability to 
detect parametric regimes where the (number of moments used in
the) present closure model is inadequate and hence appropriate
refinements (including higher order moments) are necessary.

In the projective integration method \cite{Gear01a} one takes advantage
of the slow dynamics of the coarse variables to carry out only bursts
of microscopic simulations connected via projections (in effect,
extrapolations and/or interpolations) over gaps of time.
In the same spirit of exploiting regularity, but now in space, we have
developed the so-called gaptooth scheme \cite{Kevrekidis00,Gear02c} by
evolving the full process only in an array of small spatial boxes (the
teeth) separated by empty regions (the gaps).
Clearly, the two methods are closely related by the physics of the
problem.
%
Indeed we can have a combined gaptooth-projective integration scheme;
%
this is the focus of another paper \cite{Gear02c}. 
Here, we simply want to point out the fact that in the gaptooth
method, the teeth communicate with each other via appropriate boundary
conditions for the microscopic simulations performed inside them.
And here lies the {\em raison d'etre} of this paper.

It is a well-known fact that certain features of a given equation
affect the nature of the appropriate numerical solver.
A Hamiltonian dynamics problem, for example, is best integrated by a
symplectic integrator; often, finite difference solvers of partial
differential equations (PDE) are built to respect certain properties
of the PDE, such as conservation laws.
Most importantly, the highest {\em spatial order} of an evolution
equation critically affects the types of boundary conditions leading
to a well-posed problem.

In a completely analogous manner, the way in which the microscopic
model is solved seperately in each tooth in the gaptooth scheme,
and the boundary conditions applied to the edges of each tooth,
must respect the nature of the unavailable equations and their order.
Furthermore, gaptooth algorithms compatible with conservation laws
(e.g. using fluxes to estimate temporal derivatives, see for example
\cite{EEngquist}) are predicated upon knowing that the unavailable
equation possesses certain conservation laws.

When the closed-form equation is available, some of these questions 
(e.g. the order of the highest spatial
derivative in an evolution equation) can be answered by direct
inspection.
Other issues (such as the existence of conservation laws, or integrability)
may, in the case of closed form equations, be
relatively obvious, or may require a lot of work.
%

What we explore in this paper is the development of computer-assisted
methodologies to answer the above questions when closed form equations
{\em are not available}.
The idea is that we can probe {\it the consequences} of these answers
on the dynamics of the unavailable coarse equations using microscopic,
particle- or agent-based, simulators by trying out large classes of
appropriately chosen initial conditions.
We will illustrate what we call the {\it Baby-Bathwater}
algorithm on examples of particle systems realizing the Burgers and
Korteweg-de Vries (KdV) equations, for the task of inferring the
highest order of spatial derivative on the right-hand-side, and for
answering questions concerning coarse conservation laws.

The paper is organized as follows: In Section II we briefly present
our illustrative particle-based example.
In Section III we discuss the determination of the highest order
spatial derivative active in the ``unavailable equation''.
In Section IV we explore the possible existence of conservation laws.
In the concluding Section we discuss the scope and limitations of the
procedure, as well as additional questions that may be addressed
through this approach.
An interesting ``twist'' about reverse  coarse integration arises in
discussing the exploration of possible ``coarse Hamiltonianity'' of
the unavailable coarse equation.

\section{Numerical Experiment Setup}
\label{sec:setup}

Our illustrative examples will be based on  simple numerical
experiments.
We will first, as a sanity check, demonstrate the approach using a
traditional numerical simulator of a known evolution equation as a
``black box''.
We will then substitute the simulator of the known continuum equation
with a particle-based simulator, and repeat the procedure.
Our first illustration will be the Burgers equation,
\begin{equation}
 u_t + uu_x = \nu u_{xx},
 \label{BurgersEquation}
\end{equation}
as well as a particle based simulator constructed so that the
evolution of its density resembles (at the appropriate limit,
reproduces) the Burgers evolution.
Since one of the issues to be explored is the number and type of
boundary conditions in evolving the equation, our simulations must be
possible without this {\it a priori} knowledge.
We therefore use periodic boundary conditions (PBC) enforced on
$x\in[0,2\pi)$ in all our exploratory simulations.
One of the attractive features of the Burgers is that, for any initial
profile $u(x,t=0)$, and even with PBC, the Cole-Hopf
transformation \cite{Cole51}
provides an analytical solution.
The accuracy of the numerics can thus be checked directly.

A biased random walker-based particle simulator mimicking the Burgers
dynamics was also constructed to demonstrate the direct application
of our procedure on microscopic, particle based solvers.
A detailed study of the features of this particle model is reported
elsewhere \cite{Li03}.
As a reference, the diffusion equation,
\begin{equation}
 u_t = \nu u_{xx},
 \label{DiffusionEquation}
\end{equation}
has the well-known microscopic realizations of Langevin dynamics or
unbiased random walkers. 
It is not too difficult to conjure up a similar realization motivated
by the Burgers equation using random walkers: a unit mass $\int
u(x,t)dx=1$ in the coarse description corresponds to $Z$ walkers,
where $Z$ is a large integer constant.
In the
simulation, $N$ random walkers move on $[0,2\pi)$ at discrete
timesteps $t_n=nh$. 
At each step, (a) the walkers' positions $\{x_i\}$
are sorted, (b) each walker $i$ checks out the position of the walker
$m$-places ahead, $x^{m+}_i$, and $m$-places behind, $x^{m-}_i$
(properly accounting for PBC, of course). 
The difference $x^{m+}_i-x^{m-}_i$ is inversely proportional to the
local density of walkers, therefore (c) every walker moves by $\Delta
x_i$ sampled from $N(mh/Z(x^{m+}_i-x^{m-}_i),2\nu h)$, a biased
Gaussian distribution.
The $x_i$'s are then wrapped around to $[0,2\pi)$, and the
process repeats.
This achieves a coarse-grained flux of $j\equiv u^2/2-\nu u_x$ as
motivated by the (\ref{BurgersEquation}) by assigning each walker a
drift speed of $u/2$.
Quantifying the approximation of the Burgers evolution is an
interesting subject that we take up separately in \cite{Li03}; this is
not, however, an important issue for this paper.
It is only for benchmarking purposes that the relation to a known
macroscopic equation is brought up.
One can start by presenting a microscopic evolution law, without
knowing anything about its corresponding coarse equation, and apply
our algorithms on it directly.

Relating the fine with the coarse description requires the 
use of lifting and
restriction operators\cite{Gear02a}.
Lifting constructs particle distributions conditioned on some of their
lower moment fields (here the zero-th moment, or coarse density field);
it clearly is a one-many operator, and several microscopic ``copies''
of a given macroscopic initial condition are often required as
discussed in detail in
\cite{Kevrekidis02,Gear02a}.
The restriction (here computing moment fields of a given particle
distribution) is a form of projection.
Clearly the restriction (back to coarse variables) of a lifting of
these coarse variables should be the identity (or close to it, due to
noise effects).
The lifting and restriction operators we constructed for this work,
with $u$ interpreted as the coarse density on $[0,2\pi)$ PBC, are
given in Appendix \ref{sec:density}.
Fig. \ref{reversibility} shows a result of the ``reversibility test'': we
randomly generate a coarse density $u(x)$, lift it to a random walker
distribution, then restrict back to $\tilde{u}(x)$, and observe the
very good agreement between $\tilde{u}(x)$ with $u(x)$. Notice that
the only point where this agreement may be less satisfactory is close
to local maxima or minima, where the derivative changes sign.

\begin{figure}[ht]
\centering
{\epsfig{file=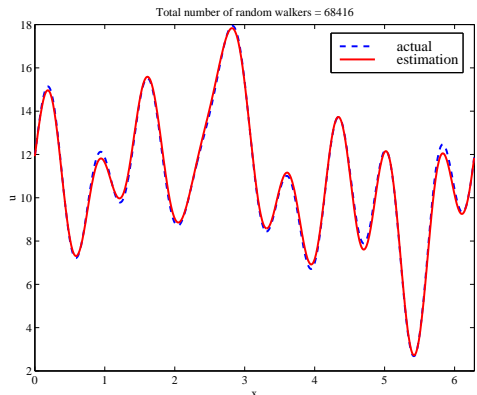, width=6.3cm,angle=0, clip=}}
\caption{Reversibility test ($\hat{\cal M}\hat{\mu}\approx \hat{I}$)
of the $\hat{\mu},\hat{\cal M}$ operators constructed in Appendix
\ref{sec:density}: $Z=1000$, $M=10$, and $u(x)$ is generated by
randomly drawing $na_n$,$nb_n$ from $N(0,1)$, $n=1..M$
(see the Appendix for details).}
\label{reversibility}
\end{figure}

\begin{figure}[ht]
\centering
{\epsfig{file=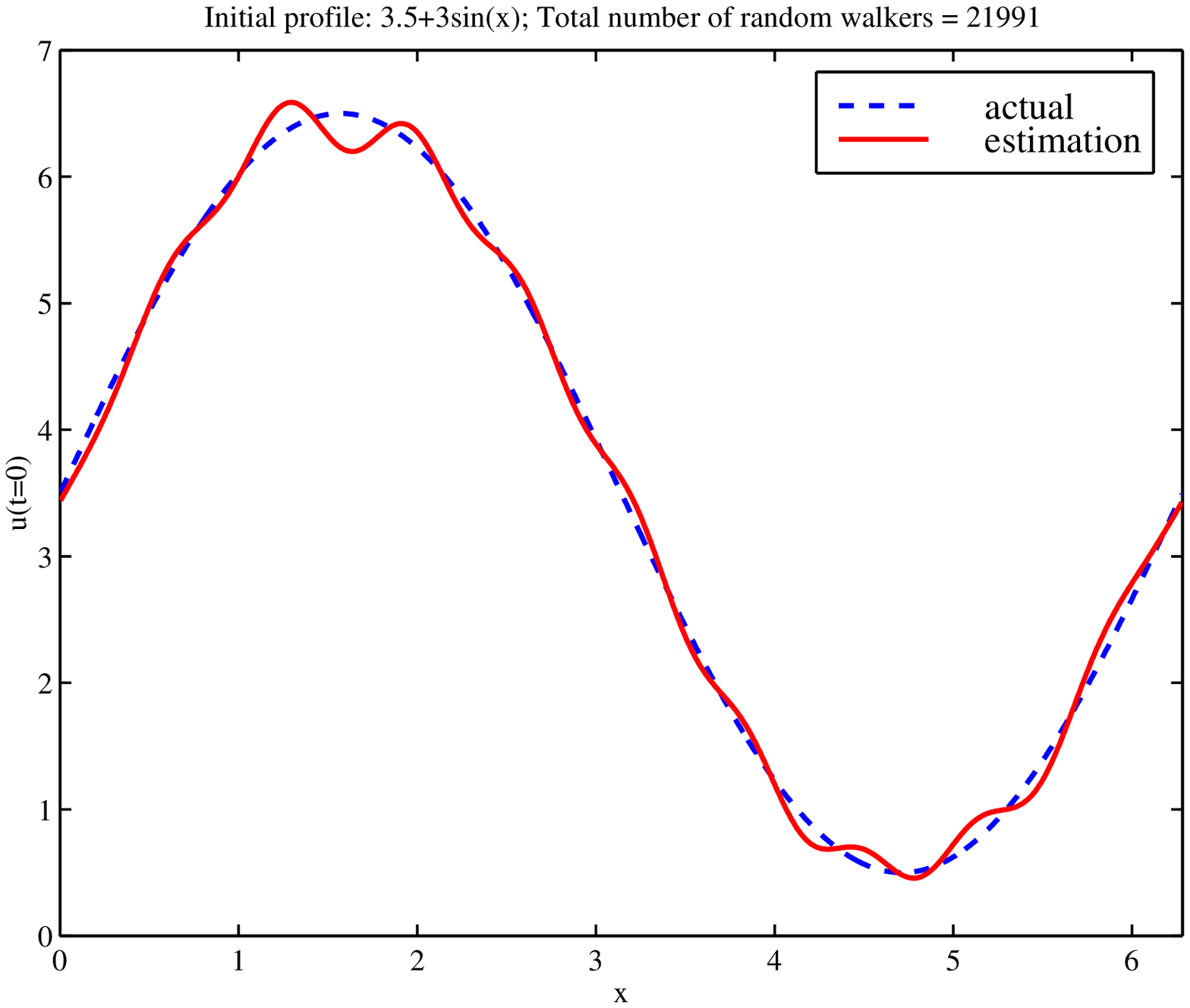, width=6.3cm,angle=0, clip=}}
(a)
\centering
{\epsfig{file=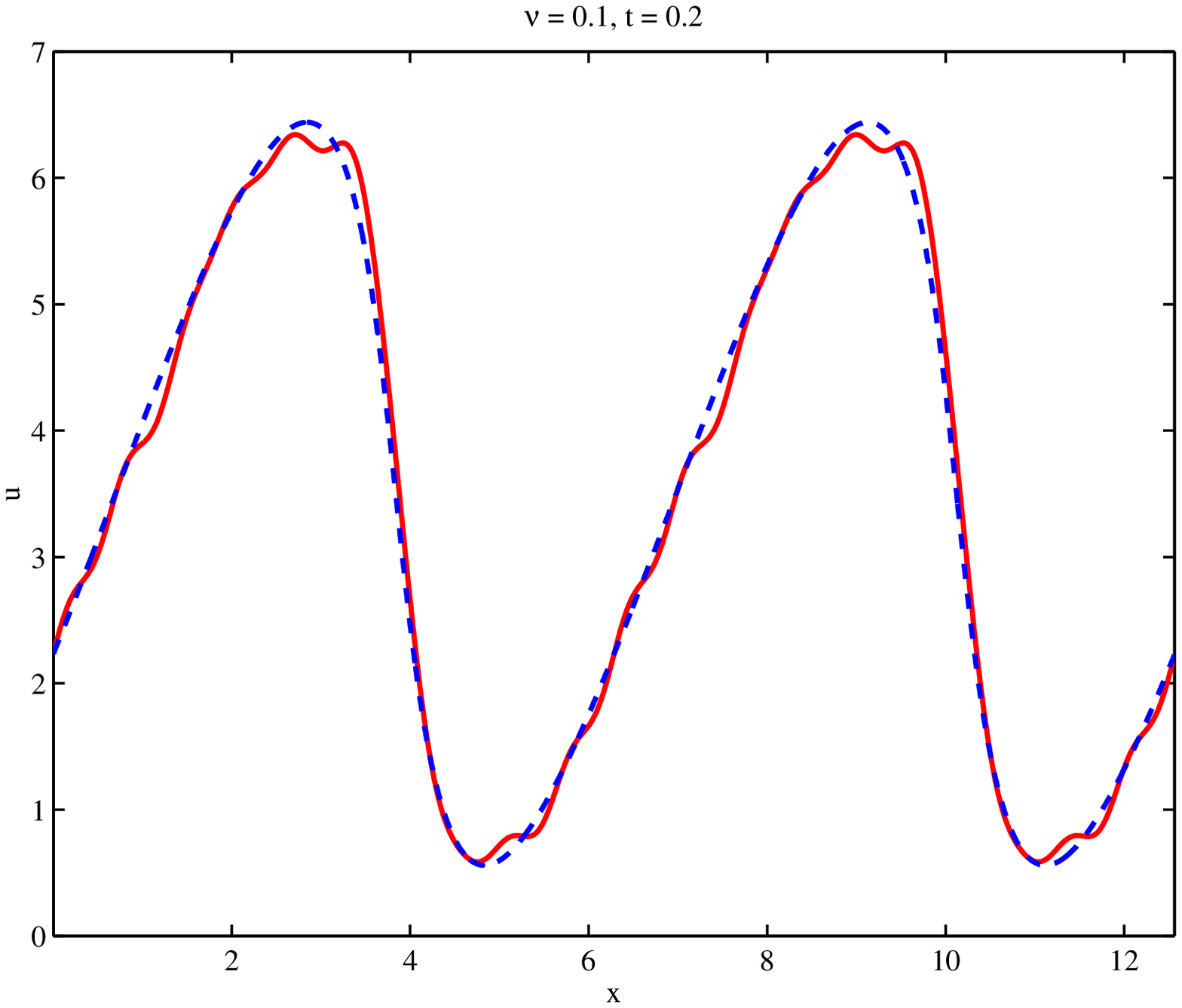, width=6.3cm,angle=0, clip=}}
(b)
\centering
{\epsfig{file=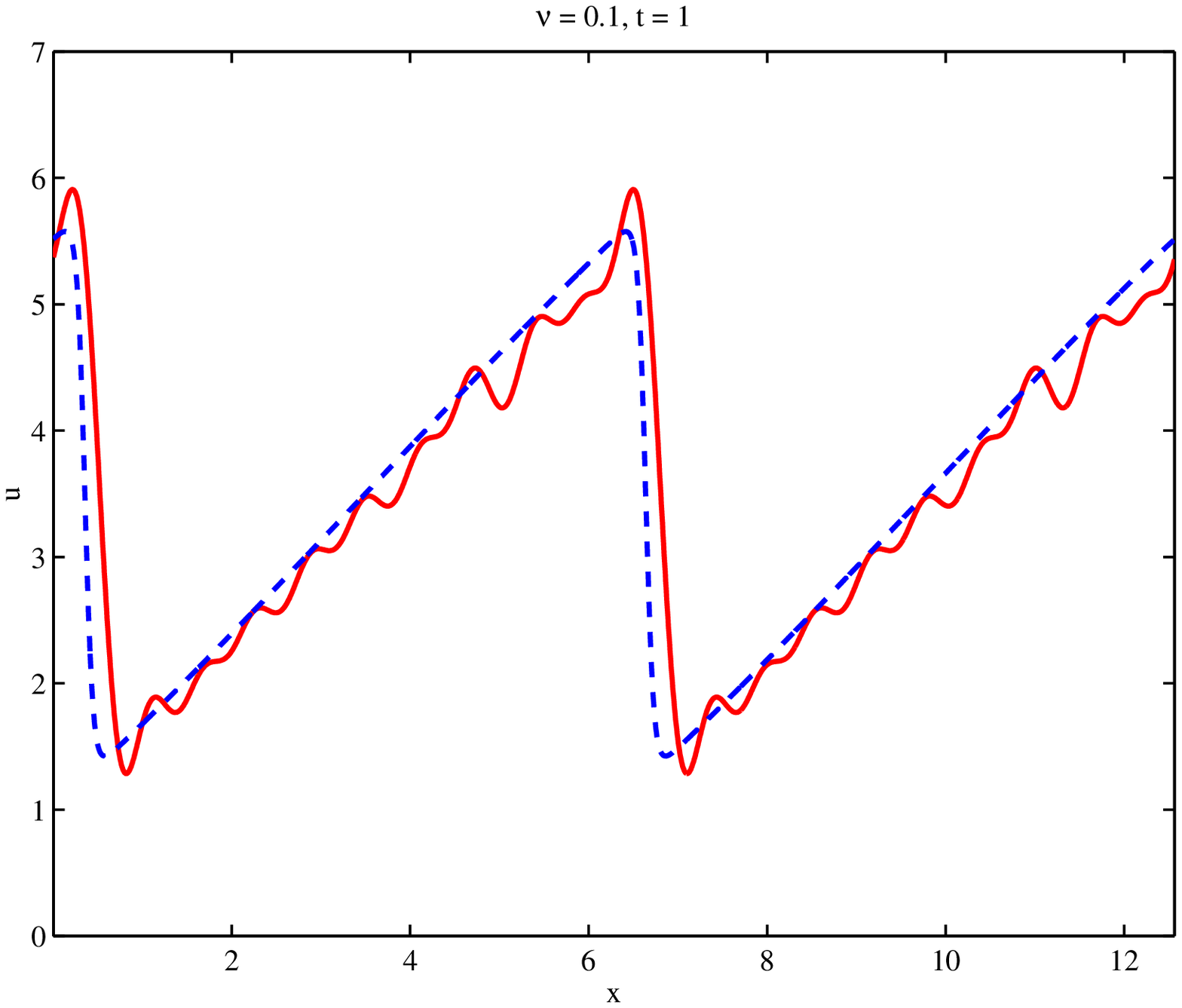, width=6.3cm,angle=0, clip=}}
(c)
\caption{Analytical solution of the Burgers at $\nu=0.1$, and
solutions of our particle-based scheme.  (a) Initial condition
$u(x,0)=3.5+3\sin(x)$, comparison between designated profile and after
$\hat{\cal M}\hat{\mu}$ for $Z=1000$, $M=10$, (b) comparison of
analytical solution and simulation (after restriction) at $t=0.2$
(tiled for ease of seeing the wave steepening), and (c) $t=1$. The
microscopic simulation is carried out with $m=10$,
$h=5\times10^{-4}$.}
\label{BurgersFullMC}
\end{figure}

While the proximity of our particle scheme to the Burgers evolution is
not the issue in this paper, we briefly illustrate the correspondence
of the evolution of an initial profile through the two approaches.
Fig. \ref{BurgersFullMC} shows the analytical solution obtained
through the Cole-Hopf transformation; it also contains the result of a
$21991$-particle simulation, after the configurations have been
processed by the $\hat{\cal M}$ operator of Appendix \ref{sec:density}
to extract the coarse density field estimate.
A small value of viscosity $\nu=0.1$ is picked to accentuate the
behavior of steepening wave-front with time.
The microscopic simulation clearly captures the important features of
the coarse behavior.
Ensemble averaging with the same initial condition in coarse field
$u(x,0)$ would reduce the error, but as we can see, even a single
microscopic simulation using a reasonable number of particles may
still perform quite efficiently.

Lastly, we mention the existence of particle methods to solve partial
differential equations (PDEs) such as the Korteweg - de Vries (KdV)
equation:
\begin{equation}
 u_t = 6uu_x - u_{xxx}
 \label{KdVEquation}
\end{equation}
which can be formulated as conservation laws
\cite{Chertock01,Chertock02}. 
In this paper we use our construction above for the Burgers example.
We should highlight once more that our ultimate purpose is {\it not} to 
construct particle
solvers of {\it given equations}, but rather to decide on features of the
{\it unavailable} coarse equations for {\it given microscopic
schemes}.

\section{Identifying The Highest Spatial Order of Coarse Variables}
\label{sec:Identification}

As we mentioned in Section \ref{sec:Introduction}, system
identification lies at the heart of the equation-free approach.
Here, we
suppose the coarse dynamics follows a certain time-evolution equation
of the form:
\begin{equation}
 u_t = f(u,u_x,u_{xx},..,u_x^{(N)}),
 \label{LocalEvolutionEquation}
\end{equation}
that is unavailable to us.
We have already identified $u$ (the ``coarse variable'' for which we
believe that a coarse deterministic equation exists in closed form), and
constructed the lifting and restriction operators that connect
macro/micro descriptions.
We seek a general approach to decide qualitative questions, such as
(a) what is $N$, and (b) whether $f$ can be written as $-\nabla\cdot
{\bf j}$, without having $f$ in closed form.
One important motivation for this lies in that ``production run''
simulations of the problem via equation-free computation (for example
through the gaptooth scheme) do not require knowledge of $f$, but are
affected by the knowledge of $N$ (through ``teeth'' boundary
conditions).
What we do have is a microscopic simulator embodied in a computer code
that can be initialized at will; the physical details of the
microscopic code are both extremely important (that is where the
``underlying physics'' lies) and - for our purposes - irrelevant: we
will use the microscopic simulation code as an ``input-output'' (I/O)
black box.
By probing the coarse I/O response of the black box, 
the question that we would like to address is whether we can decide 
on (a) and/or (b).

It may appear initially that we are trying to answer a circular
question: in order to probe the coarse input-output response of a
microscopic simulator we need to run it, and in order to run it we
need well-posed boundary conditions, which - among other factors -
depend on (a) and (b).
To cut the knot, we use (for the decision stage exploratory runs)
the Born-von Karman periodic boundary conditions.
We are going to assume that the microscopic
simulations can be carried out in PBC, which is an option prevalent
among microscopic simulators.
This enables us to probe the
system's response to only the initial $u(x,0)$ profile input.

The so-called baby-bathwater identification scheme works as follows:
\renewcommand{\labelenumi}{{(\roman{enumi})}}
\begin{enumerate}
\item Take an integer $n$, starting from $1$.

\item Pick a random point $x_0$ in the spatial periodic box.

\item Generate $n$ random numbers, designated as
$u(x_0,0)$, $u_x(x_0,0)$, $u_{xx}(x_0,0)$, .., $u_x^{(n-1)}(x_0,0)$ of
$u(x,0)$.

\item Generate a conditionally random profile $u(x,0)$ compatible with
the PBC and consistent with  the above $u(x_0,0)$, $u_x(x_0,0)$,
$u_{xx}(x_0,0)$, .., $u_x^{(n-1)}(x_0,0)$ requirements. 
This can almost always be accomplished, for example, by summing $2L$
sine and cosine harmonics of the PBC:
\begin{equation}
 u(x,0) = b_0 + \sum_{i=1}^L a_l \sin(lx) + b_l\cos(lx), \;\; x\in [0,
 2\pi), \label{InitialRandomProfile}
\end{equation}
with $L>\lceil n/2\rceil$. 
Because we have $2L+1$ coefficients, even though there are $n$
constraints to satisfy, we still have some random degrees of freedom
left in (\ref{InitialRandomProfile}).
In practice, this initialization can be accomplished by applying
conjugate gradient minimization of the $n$-dimensional residual norm
starting from a random $\{a_l,b_l\}$ vector.

\item Lift $u(x,0)$ of (\ref{InitialRandomProfile}), run it in the
microscopic simulator for time $\Delta$, restrict it back to
$\tilde{u}(x,\Delta)$, and estimate:
\begin{equation}
 \tilde{u}_t(x_0,0) \equiv
 \frac{\tilde{u}(x_0,\Delta)-\tilde{u}(x_0,0)}{\Delta}.
 \label{MicroscopicUtEstimation}
\end{equation}
Note that here $\tilde{u}(x_0,0)$ instead of $u(x_0,0)$ is used in the
finite difference. This will cancel some internal noise from the
lifting and restriction operations.

\item Repeat step (v) $I$ times to obtain an ensemble averaged
$\tilde{u}_t(x_0,0)$ to reduce the microscopic noise.

\item Repeat step (iv) $J$ times, collect the $\tilde{u}_t(x_0,0)$
estimates:
\begin{equation}
  ( \tilde{u}_t^1(x_0,0), \tilde{u}_t^2(x_0,0), ...,
 \tilde{u}_t^J(x_0,0) ), \label{UtEstimates}
\end{equation}
compute the sample variance $\sigma^2(\tilde{u}_t(x_0,0))$.

\item Repeat step (ii) $K$ times, compute the averaged sample variance
$\langle\sigma^2(\tilde{u}_t)\rangle_n$.

\item Go back to step (i), $n\rightarrow n+1$. $N$ is identified when
from $n=N$ to $n=N+1$, the averaged sample variance
$\langle\sigma^2(\tilde{u}_t)\rangle_{N+1}$ decreases drastically to
practically $0$.
\end{enumerate}

\begin{figure}[ht]
\centering
{\epsfig{file=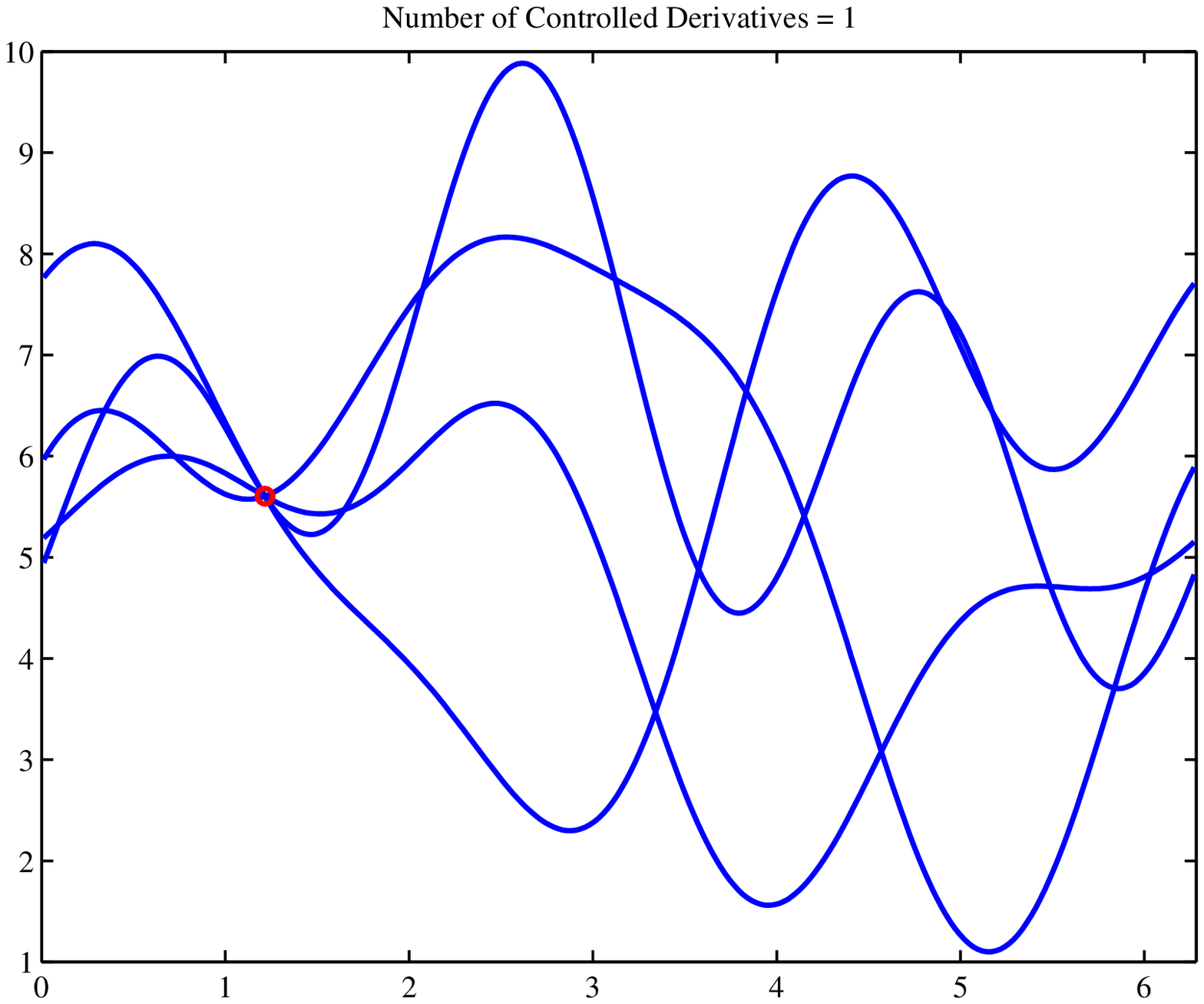, width=6.3cm,angle=0, clip=}
(a)
\epsfig{file=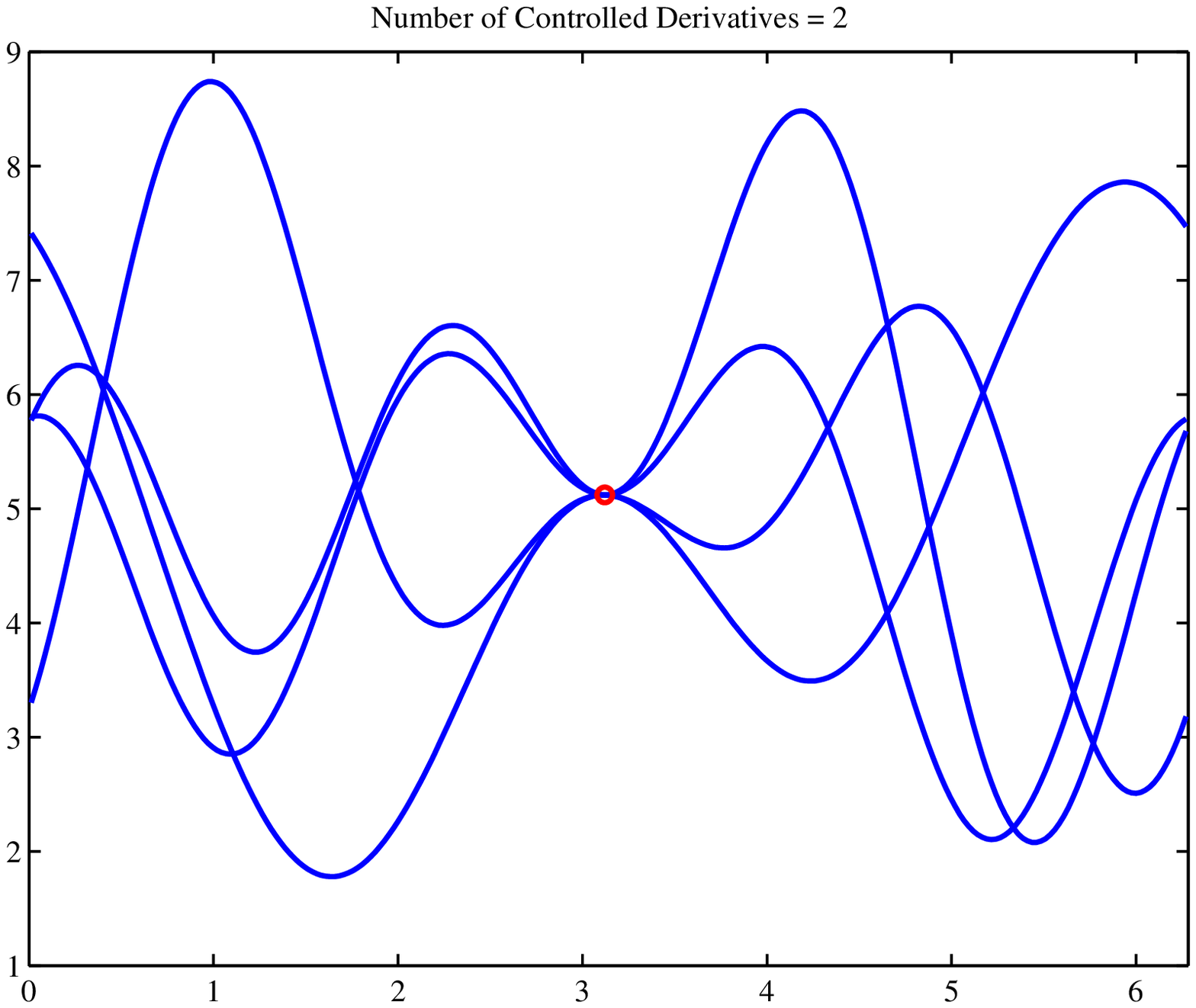, width=6.3cm,angle=0, clip=}
}
 (b)
\centering
{\epsfig{file=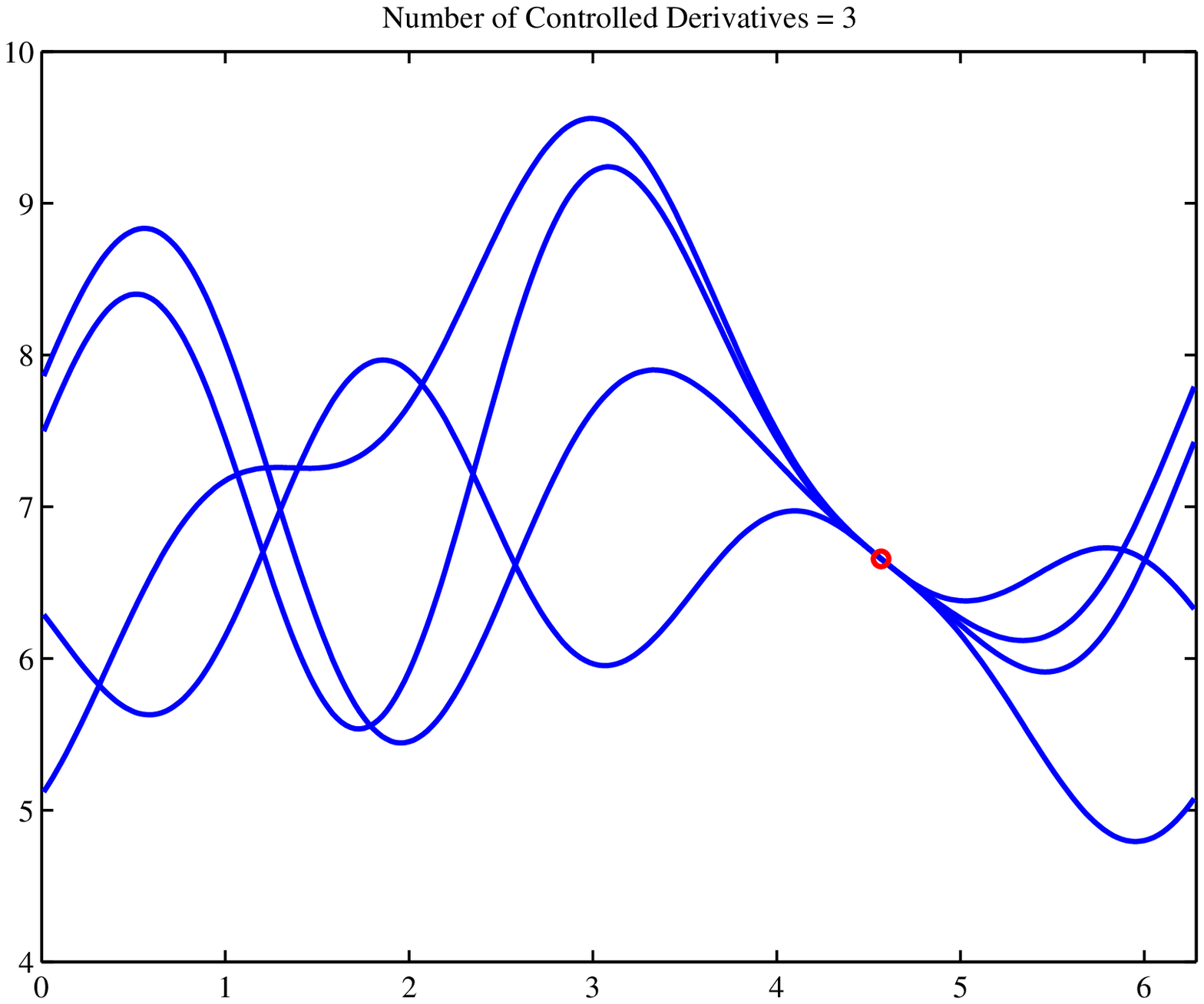, width=6.3cm,angle=0, clip=}
(c)
\epsfig{file=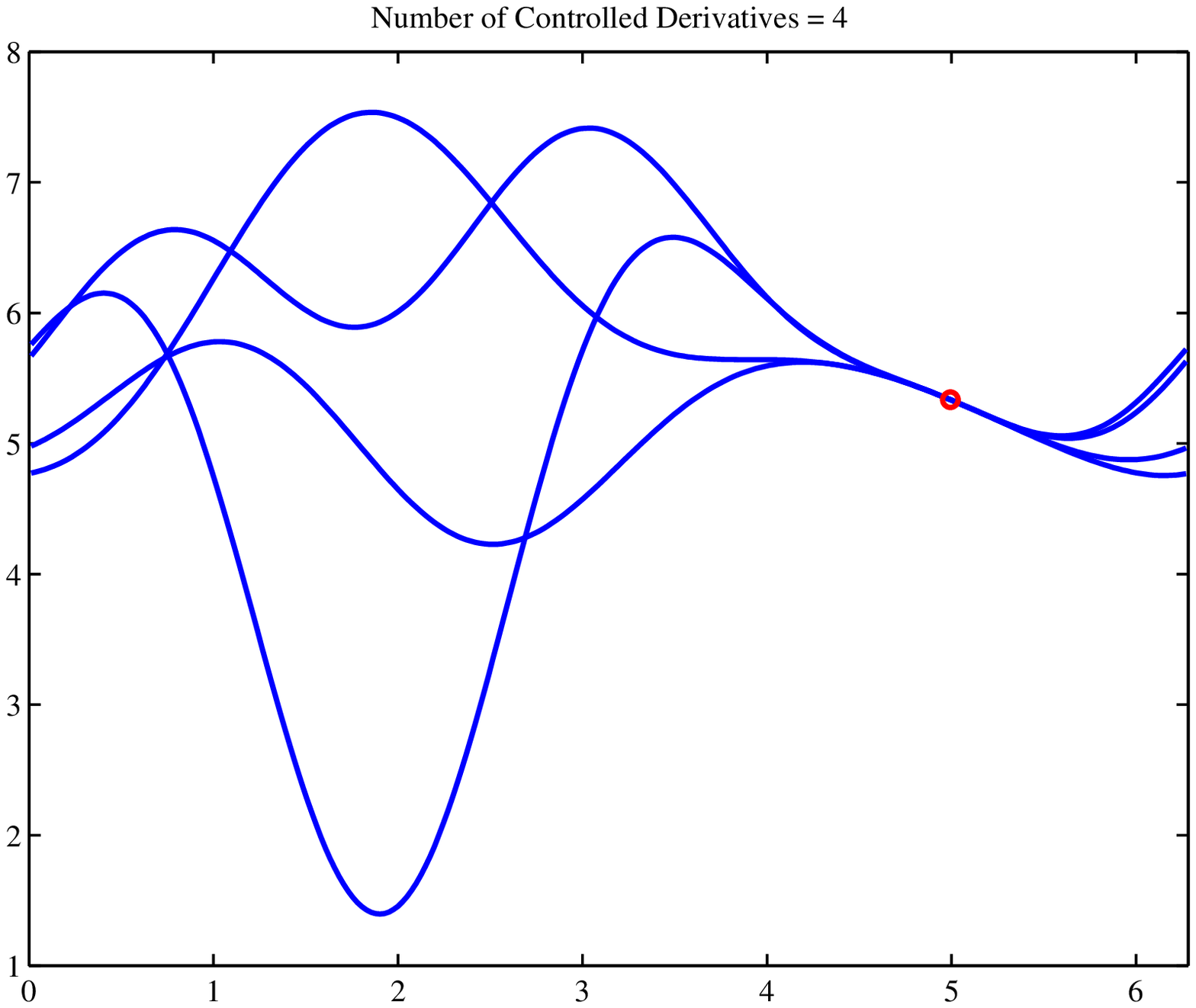, width=6.3cm,angle=0, clip=}
}
(d)
\caption{Families of random initial profiles $u(x,0)$ ($J=4$). (a)
$n=1$, (b) $n=2$, (c) $n=3$, (d) $n=4$ controlled initial
derivatives. To avoid confusion notice that control of $n=1$
derivatives means that only $u(x_0,0)$ is identical between
the runs, $n=2$ means that $u(x_0,0)$ and $u_x(x_0,0)$ are
identical and so on.}
\label{BurgersPDEIdentificationCtrl}
\end{figure}

Fig. \ref{BurgersPDEIdentificationCtrl} shows such families of
constructed initial profiles with progressively more controlled
initial derivatives.  
The basic idea is very simple: even though $f$ could have complicated
functional dependencies on $u,u_x,u_{xx},..,u_x^{(N)}$, if they are
all fixed, $u_t$ should have no dispersion even as
$u_x^{(N+1)},u_x^{(N+2)},...$ are varied randomly.
The ``critical integer order'' $N$ is identified when the variance at
$N$ controlled derivatives $u(x_0,0)$, $u_x(x_0,0)$, $u_{xx}(x_0,0)$,
.., $u_x^{(N-1)}(x_0,0)$ jumps to a finite value; we then have already
thrown out the ``baby'' (the highest relevant spatial derivative
$u_x^{(N)}$) with the ``bathwater'' (the higher, non-relevant ones).

It is important to recognize that the time derivative estimation
(\ref{MicroscopicUtEstimation}) does not occur instantaneously.
A short ``healing'' period should elapse, during which the higher
cumulants of the lifted phase space (micro-state) distribution become
functionals of the lower order, slow governing cumulants.
This separation of time scales, which fundamentally underlies the
existence of a deterministic coarse equation closing with the lower
cumulants, is discussed in more detail in \cite{Kevrekidis02}.

\begin{figure}[ht]
\centering
{\epsfig{file=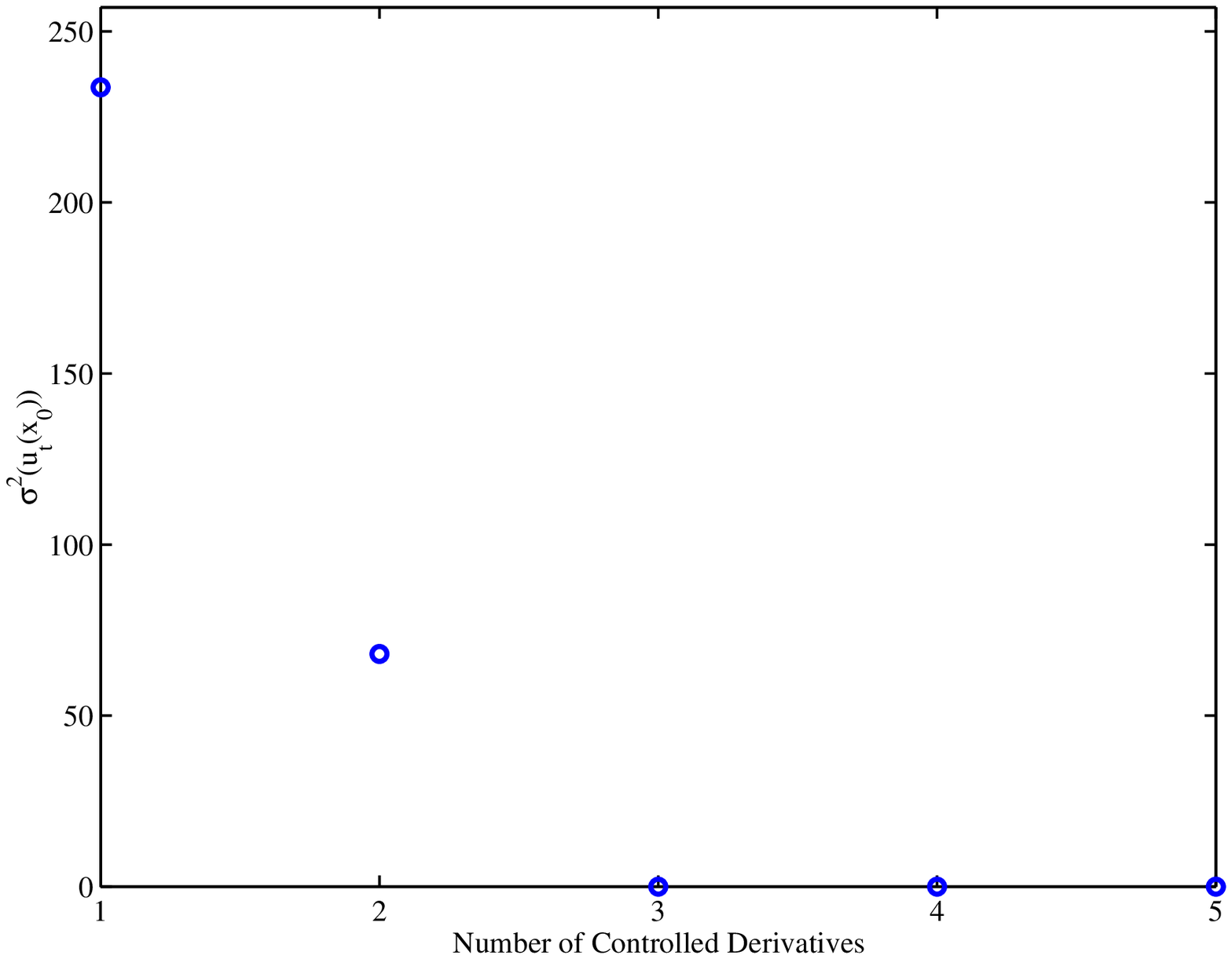, width=6.3cm,angle=0, clip=}
(a)
\epsfig{file=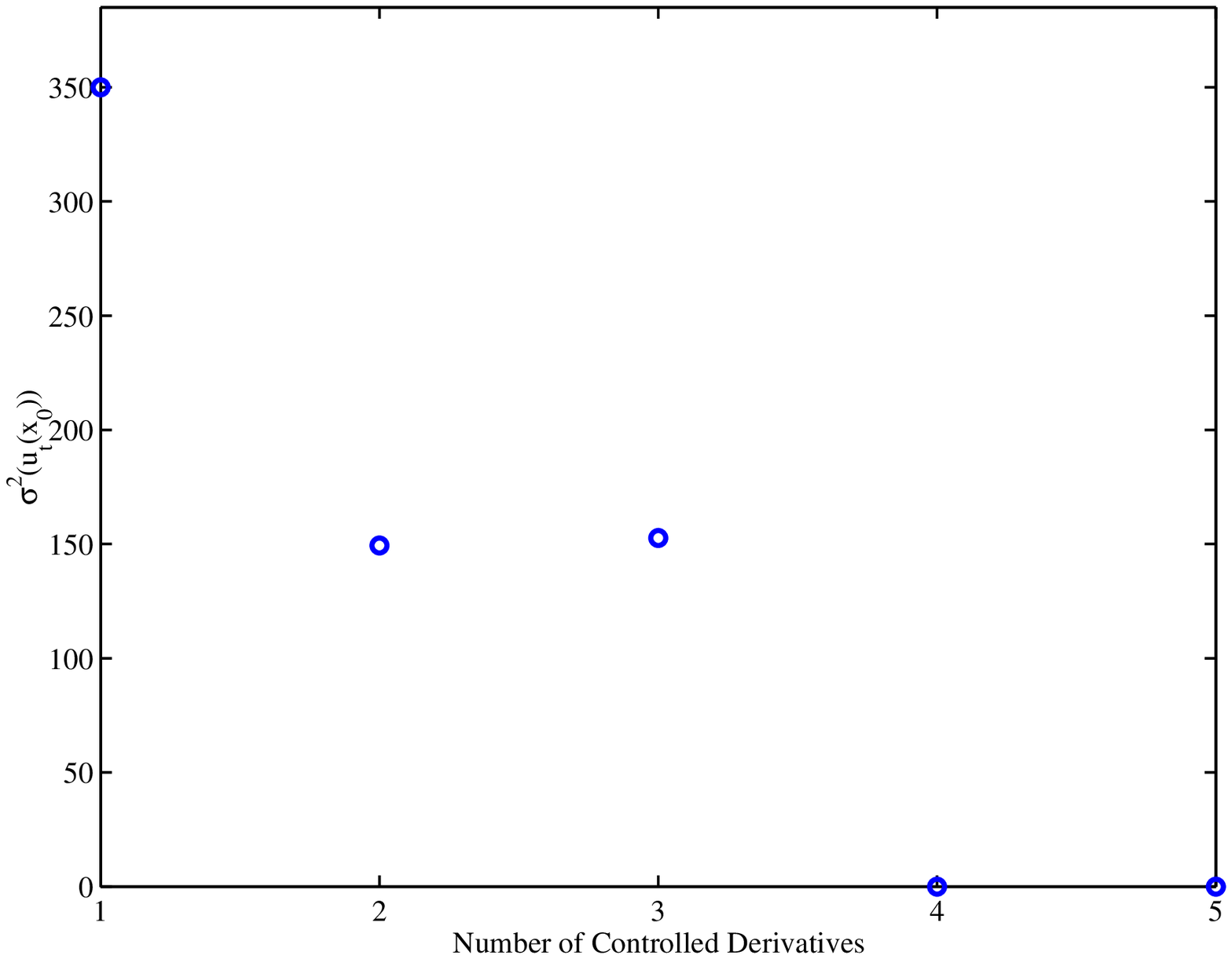, width=6.3cm,angle=0, clip=}
}
(b)
\caption{(a) Identification the order of the highest (spatial) derivative in
 the Burgers finite-difference PDE
time-stepper (\ref{BurgersFiniteDifferenceIntegrator}). (b)
Identification of the order of the highest (spatial) derivative for
 the KdV finite-difference PDE time-stepper
(\ref{KdVFiniteDifferenceIntegrator}).}
\label{PDEIdentification}
\end{figure}

As a sanity check, this algorithm is also applied to a traditional
continuum PDE time-stepper ``black box'' first.
Fig. \ref{PDEIdentification}(a) and \ref{PDEIdentification}(b) show
the results of applying our decision scheme to forward Euler
finite-difference PDE solvers of the Burgers and KdV equations,
respectively.
A spatial mesh of
$\Delta x=2\pi/100$ is adopted, and we define
\begin{eqnarray}
 u_x^{mk} \equiv && \frac{u_{(m+1)\Delta x,kh}-
 u_{(m-1)\Delta x,kh}}{2\Delta x},  \\
 u_{xx}^{mk} \equiv &&
 \frac{u_{(m+1)\Delta x,kh}+u_{(m-1)\Delta x,kh}-2u_{m\Delta x,kh}}
 {\Delta x^2}, \\
 u_{xxx}^{mk} \equiv &&
 \frac{u_x^{m+1,k}+u_x^{m-1,k}-2u_x^{mk}}{\Delta x^2}, \\
 u_{\rm avg}^{mk} \equiv && \frac{u_{(m+1)\Delta x,kh}+u_{m\Delta x,kh}
 +u_{(m-1)\Delta x,kh}}{3}.
 \label{CentralDifferences}
\end{eqnarray}
We use,
\begin{equation}
 \frac{u_{m\Delta x,(k+1)h}-u_{m\Delta x,kh}}{h} = 
 \nu u_{xx}^{mk} - u_{m\Delta x,kh} u_x^{mk},
 \label{BurgersFiniteDifferenceIntegrator}
\end{equation}
to integrate the Burgers equation forward, and
\begin{equation}
 \frac{u_{m\Delta x,(k+1)h}-u_{m\Delta x,kh}}{h} = 
 6 u_{\rm avg}^{mk} u_x^{mk} - u_{xxx}^{mk},
 \label{KdVFiniteDifferenceIntegrator}
\end{equation}
to integrate the KdV equation forward. 
$u(x_0,kh)$ is obtained by cubic spline over $\{u_{m\Delta x,kh}\}$,
and $u_t(x_0,0)$ is evaluated by finite differences, same as in
(\ref{MicroscopicUtEstimation}).
As can be seen in Fig. \ref{PDEIdentification}(a) and
\ref{PDEIdentification}(b), $N$ is identified to be $2$ using the
Burgers PDE time-stepper and $3$ using the KdV PDE time-stepper: the
variances drop by more than four decades in both cases when going from
$N$ to $N+1$ controlled derivatives.
To see where the remaining ``noise'' comes from, note that,
\begin{equation}
 u(x_0,\Delta)-u(x_0,0) \;=\; u_t(x_0,0) \Delta + u_{tt}(x_0,0)
 \frac{\Delta^2}{2} + ...,
\end{equation}
and clearly $u_{tt}(x,0)$ has higher-than-$u_x^{(N)}$ spatial
derivative dependencies, which are, however, scaled by $\Delta$
compared to the leading term.
Thus the sample variances should drop by $\sim\Delta^2$ for $n>N$,
which explains the observed magnitude of the four-decade decrease.

\begin{figure}[ht]
\centering
{\epsfig{file=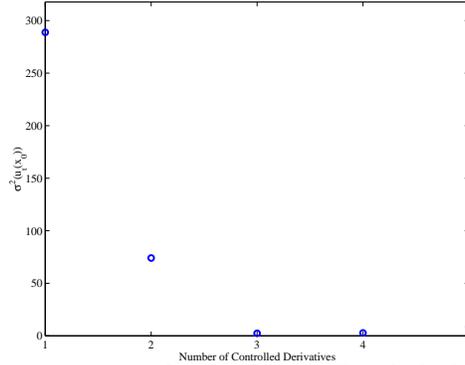, width=6.3cm,angle=0, clip=}}
\caption{Identification of the Burgers microscopic simulator of
Section \ref{sec:setup} with the lifting and restriction operators of
Appendix \ref{sec:density}. Here, $\nu=1$, $Z=10000$, $m=100$, $I=10$,
$\Delta=0.01$.}
\label{BurgersIdentification}
\end{figure}

We then apply the identification scheme to the microscopic simulator
of Section \ref{sec:setup}, with the lifting and restriction operators
constructed in Appendix \ref{sec:density}.
The results are shown in Fig. \ref{BurgersIdentification}.
Under favorable conditions such as $\nu=1$ and $m=100$, it takes about
$10$ minutes of computer time on a single 1GHz-CPU personal computer
to obtain a reasonably good microscopic noise reduction so the
variance drops by about $2$ decades going from $n=2$ to $n=3$.
Under unfavorable conditions such as $\nu=0.1$ and $m=1$, it can take
up to $1,000$ minutes of computer time to obtain the same $2$ decades
drop.
Compared with the deterministic finite-difference PDE time-steppers,
identification of a microscopic simulator is undoubtedly much more
computationally intensive, even though fundamentally there is no
difference between the two ``black boxes''.
The problem of microscopic noise reduction is a persistent issue among
all ``equation-free'' methods including
bifurcation \cite{Kevrekidis02}, projective / gaptooth
integration \cite{Kevrekidis00,Gear02c}, and identification, and calls
for a unified treatment.
This ``one time'' decision, performed at the beginning of studying
a problem, will critically affect
subsequent production runs of the microscopic simulator.
%

Here, one must pay special attention to the rank ($M$) of the
restriction operator (see Appendix \ref{sec:density}).
As can be seen
in Figs. \ref{reversibility} and \ref{BurgersFullMC}, our proposed
restriction operator satisfies the constraint of  reversibility and
also accurately represents 
the profile's
long-time evolution.
%
However, these merits do not guarantee
{\em automatically} good short-time $\tilde{u}_t$ estimates by
finite-difference. 
Special attention must be paid to the restriction
operator $\hat{\cal M}$: for example, if the highest harmonic in
(\ref{InitialRandomProfile}) for $u(x,0)$ is $L$, then with $M=L$, we
can get good reversibility test of $u(x,0)$. 
Unless we use $M=2L$ for restricting the Burgers microscopic dynamics,
however, we would {\em not} get a good estimate of $u_t$, because the
nonlinear interaction $uu_x$ in (\ref{BurgersEquation}) creates higher
harmonics in $u_t$ up to $2L$.
If $M=L$ is still used, it is equivalent to forcing a least-square
projection of a $4L+1$ vector to a $2L+1$ subspace, which may work
well enough in the long term, but is too inaccurate for short-term
finite difference estimates.
Unless this is taken care of, the $u_t$ estimate using our $\hat{\cal
M}$ is found to not even be superior to a crude bin-count density
estimator with bin-width $(2\pi/n)/8$ about $x_0$, as $2\pi/n$ is the
shortest wavelength in $u(x,0)$.

Lastly, we note that (\ref{LocalEvolutionEquation}) represents a wide
category of coarse dynamics; those with higher time-derivatives and
mixed derivatives can be converted to a multi-variate version of
(\ref{LocalEvolutionEquation}) and the baby-bathwater identification
scheme will still, in principle, work.
A notable exception is the incompressible fluid dynamics case, where
the sound-speed is infinite and the pressure plays the role of a
global Lagrange multiplier.
The incompressible fluid model is but a mathematical idealization of a
certain physical limit.
It is nonetheless useful and important enough, that the fact that it
is not directly amenable to the baby-bathwater identification is worth
mentioning.
In general, the baby-bathwater identification presented here will not
work for dynamics with instantaneous remote-action over macroscopic
lengthscales, such as,
\begin{equation}
 u_t(x,t) = \int d\xi u(\xi,t) K(x-\xi),
\end{equation}
for which it is easy to show that $u_t(x,t)$ correlates with infinite
number of local spatial derivatives $\{u_x^{(n)}(x,t)\}$.  %

\section{Identifying Conservation Laws}
\label{sec:Conservation}

In section \ref{sec:Identification} above we address the concern of
how to identify the highest spatial derivative of an unavailable
coarse equation of the type (\ref{LocalEvolutionEquation}).
It is natural to try to decide other qualitative questions: for
example, whether the coarse dynamics conserve a specific quantity,
\begin{equation}
 G \equiv \int g(u,u_x,u_{xx},..,u_x^{(N^{\prime\prime})}) dx,
\end{equation}
or not. 
In the simplest case, we ask whether $g \equiv u$ is conserved.
We note that is equivalent to asking whether the RHS of
(\ref{LocalEvolutionEquation}) can be written as,
\begin{equation}
 f(u,u_x,u_{xx},..,u_x^{(N)}) = -\partial_x
 j(u,u_x,u_{xx},..,u_x^{(N^\prime)}),
 \label{FluxExistence}
\end{equation}
or not. 
Alternatively, we ask whether there exists
$j(u,u_x,u_{xx},..,u_x^{(N^\prime)})$ such that,
\begin{equation}
 \frac{d}{dt} \int_{x_0}^{x_1} u(x,t) dx = j(x_0,t) - j(x_1,t),
 \label{FluxMassIntegralRelation}
\end{equation}
for arbitrary $x_0$,$x_1$. 
Whereas in section \ref{sec:Identification} we try to identify
features of $f(u,u_x,u_{xx},..,u_x^{(N)})$ through
(\ref{LocalEvolutionEquation}), here we can try to identify
consequences of $j(u,u_x,u_{xx},..,u_x^{(N^\prime)})$ and its features
through (\ref{FluxMassIntegralRelation}).
The process of the baby-bathwater identification can be carried over;
the only difference is that it is going to be a
{\em boundary} scheme.
In one dimension, the boundary sheme reduces to a  {\em two}-point scheme
as follows:
\renewcommand{\labelenumi}{{(\roman{enumi})}}
\begin{enumerate}
\item Take an integer $n$, starting from $1$.

\item Pick two random points $x_0$ and $x_1$ in the PBC.

\item Generate $2n$ random numbers, which are to be designated
$u(x_0,0)$, $u_x(x_0,0)$, $u_{xx}(x_0,0)$, .., $u_x^{(n-1)}(x_0,0)$
and $u(x_1,0)$, $u_x(x_1,0)$, $u_{xx}(x_1,0)$, ..,
$u_x^{(n-1)}(x_1,0)$, of $u(x,0)$.

\item Generate a conditionally random profile $u(x,0)$ compatible with
the PBC that is consistent with the above $u(x_0,0)$, $u_x(x_0,0)$,
$u_{xx}(x_0,0)$, .., $u_x^{(n-1)}(x_0,0)$ and $u(x_1,0)$,
$u_x(x_1,0)$, $u_{xx}(x_1,0)$, .., $u_x^{(n-1)}(x_1,0)$ requirements.
This can always be done by (\ref{InitialRandomProfile})
with $L>n$. 
As we discussed above,  we have $2L+1$ coefficients, even though there are
$2n$ constraints to satisfy, we still have some random degrees of
freedom left in $u(x,0)$.

\item Lift $u(x,0)$ of (\ref{InitialRandomProfile}), run it in the
microscopic simulator for time $\Delta$, restrict it back to
$\tilde{u}(x,\Delta)$, estimate:
\begin{equation}
 \tilde{U}_t(0) \equiv \frac{\int_{x_0}^{x_1}
 \tilde{u}(x^\prime,\Delta) dx^\prime -\int_{x_0}^{x_1}
 \tilde{u}(x^\prime,0) dx^\prime}{\Delta}.
 \label{ConservationMicroscopicUtEstimation}
\end{equation}

\item Repeat step (v) $I$ times to obtain an ensemble averaged
$\tilde{U}_t(0)$ to reduce the microscopic noise.

\item Repeat step (iv) $J$ times, collect the $\tilde{U}_t(0)$
estimates:
\begin{equation}
  ( \tilde{U}_t^1(0), \tilde{U}_t^2(0), ..., \tilde{U}_t^J(0) ),
\end{equation}
compute the sample variance $\sigma^2(\tilde{U}_t(0))$.

\item Repeat step (ii) $K$ times, compute the averaged sample variance
$\langle\sigma^2(\tilde{U}_t)\rangle_n$.

\item Go back to step (i), $n\rightarrow n+1$. A conservation law is
positively identified when going from $n=N^\prime$ to $n=N^\prime+1$,
the averaged sample variance
$\langle\sigma^2(\tilde{U}_t)\rangle_{N^\prime+1}$ decreases
drastically to practically $0$.
\end{enumerate}

\begin{figure}[ht]
\centering 
{\epsfig{file=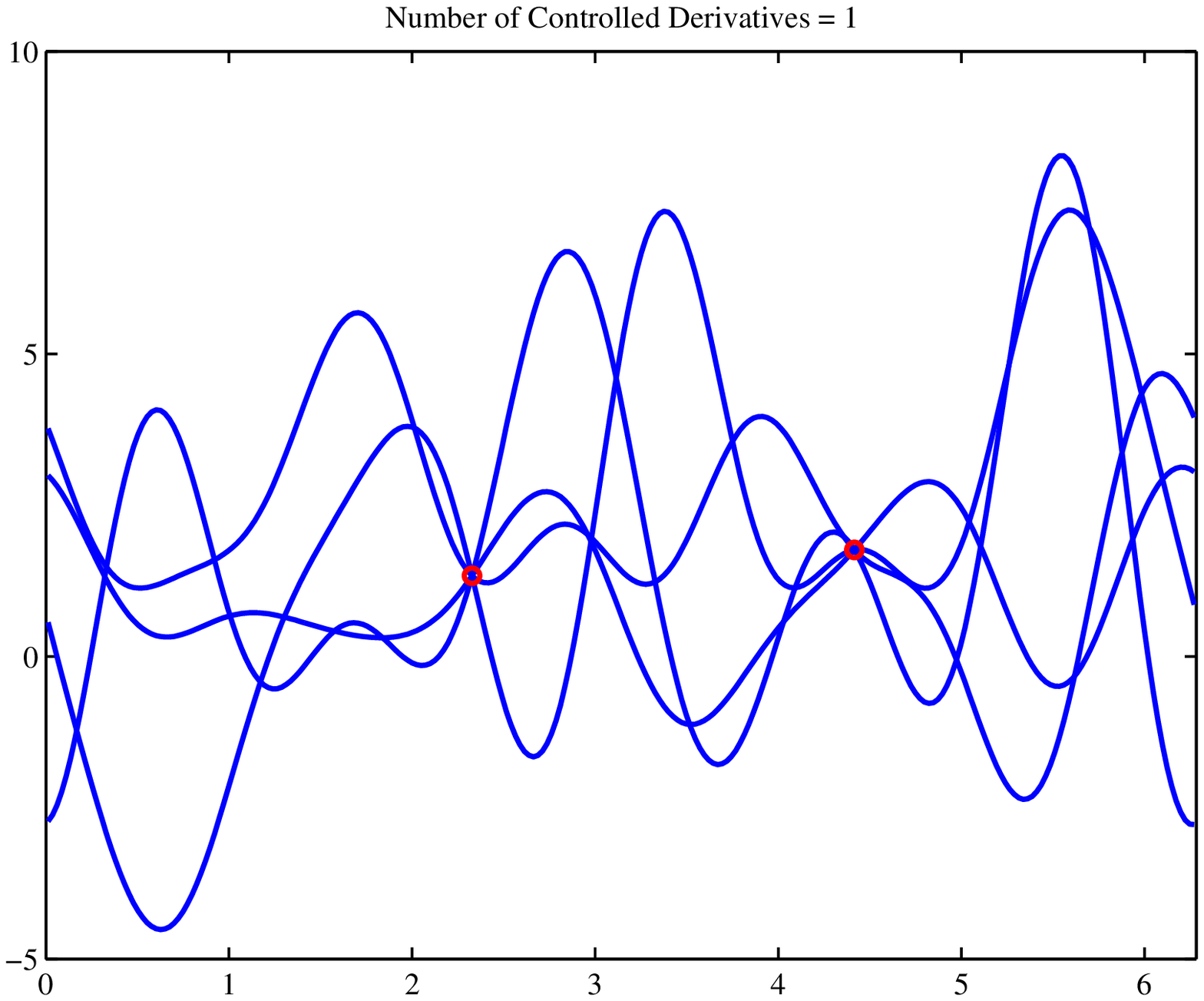, width=6.3cm,angle=0, clip=}
(a)
\epsfig{file=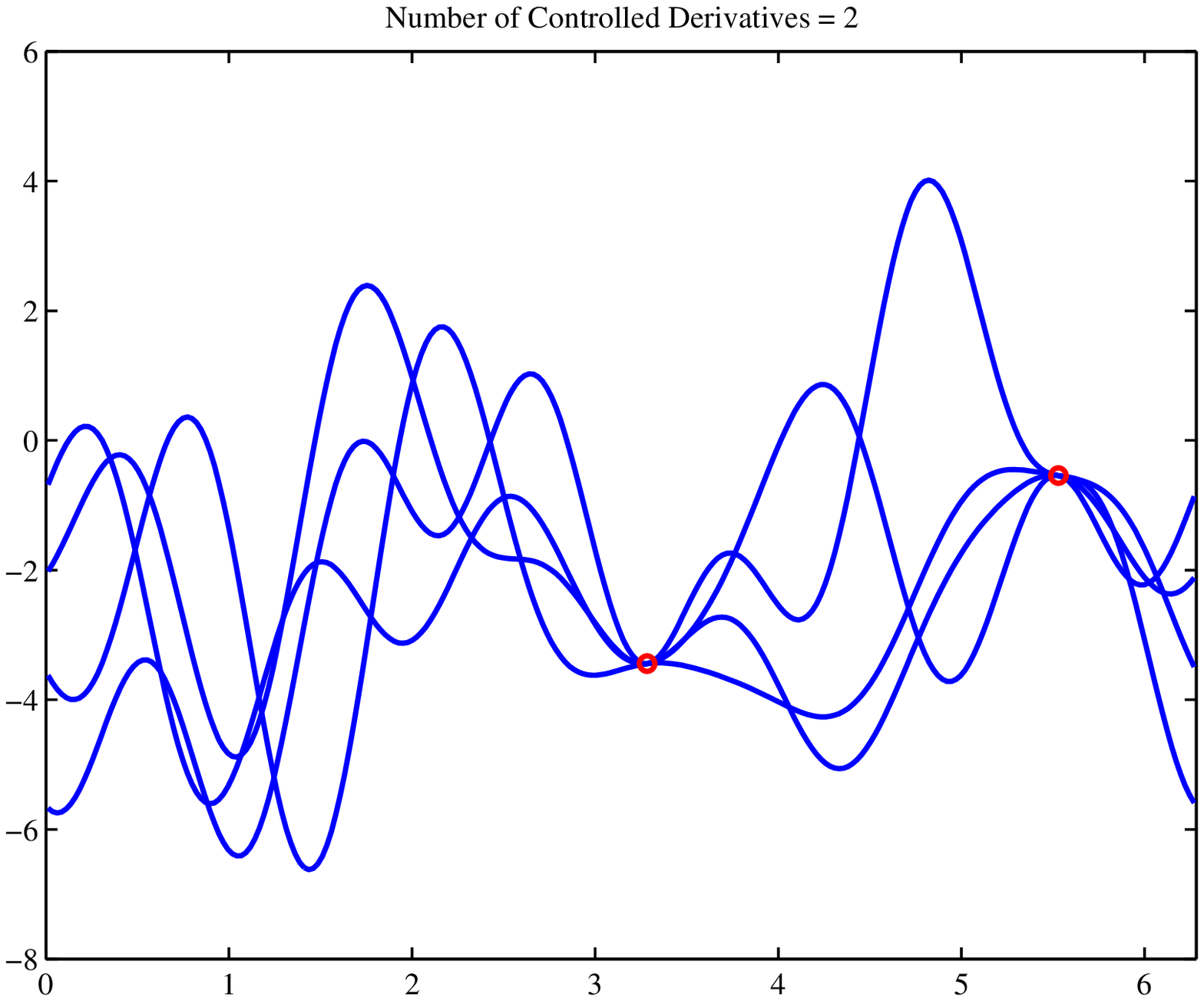, width=6.3cm,angle=0, clip=}
}
(b) 
\centering
{\epsfig{file=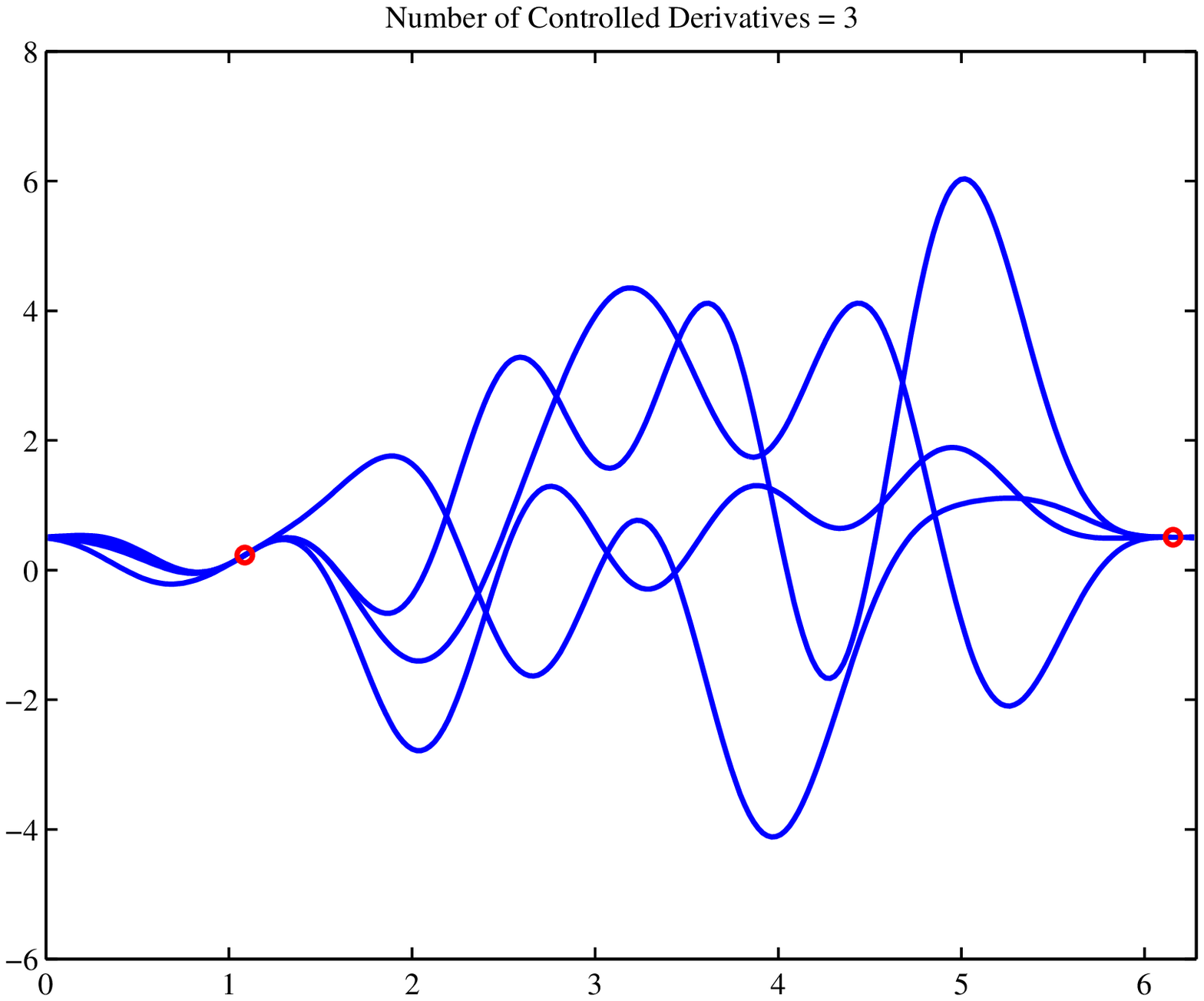, width=6.3cm,angle=0, clip=}
(c)
\epsfig{file=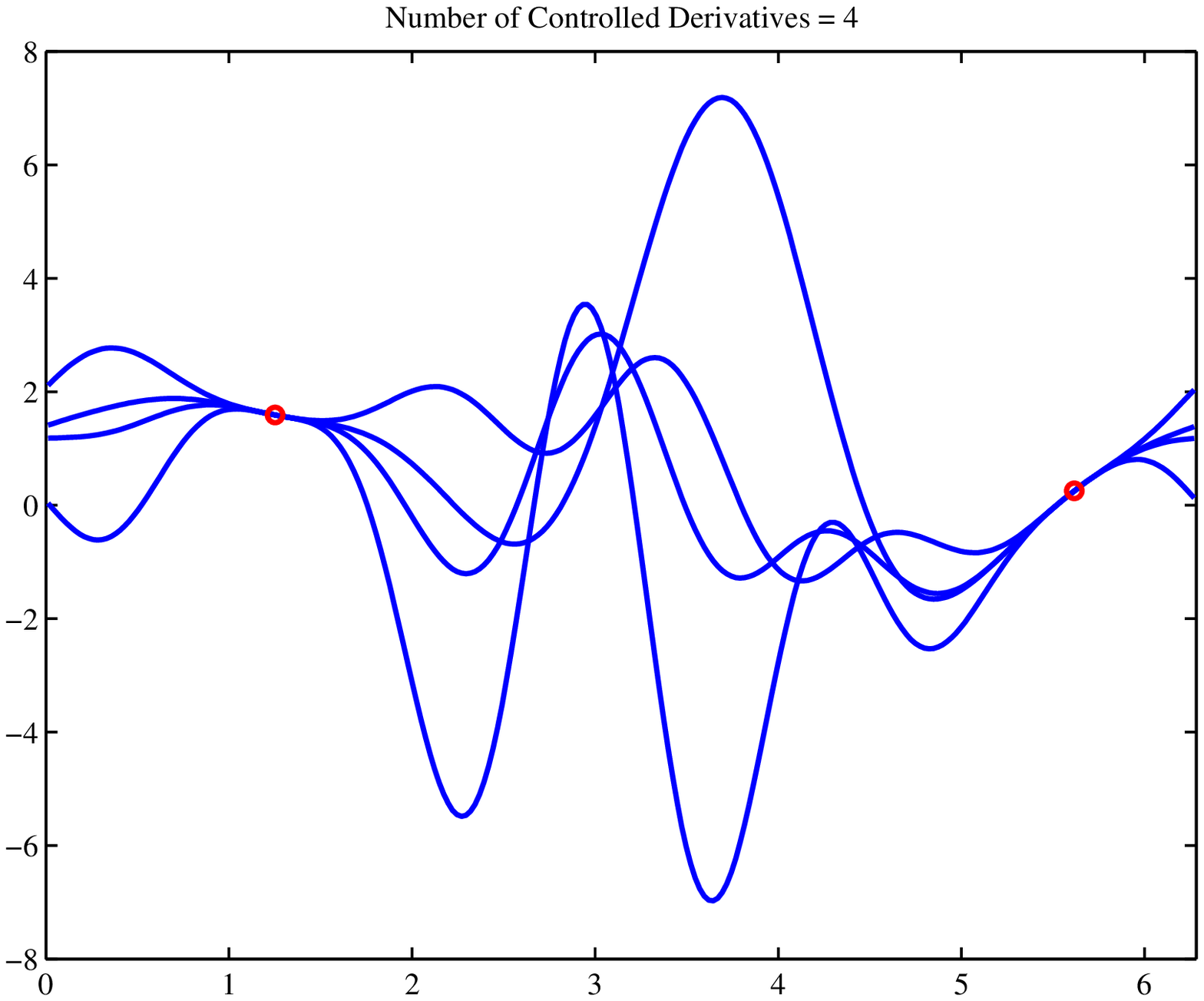, width=6.3cm,angle=0, clip=}
} (d)
\caption{Families of random initial profiles $u(x,0)$ ($J=4$) for
conservation law identification. (a) $n=1$, (b) $n=2$, (c) $n=3$, (d)
$n=4$ controlled initial derivatives.}
\label{BurgersPDEConservationCtrl}
\end{figure}

\begin{figure}[ht]
\centerline
{\epsfig{file=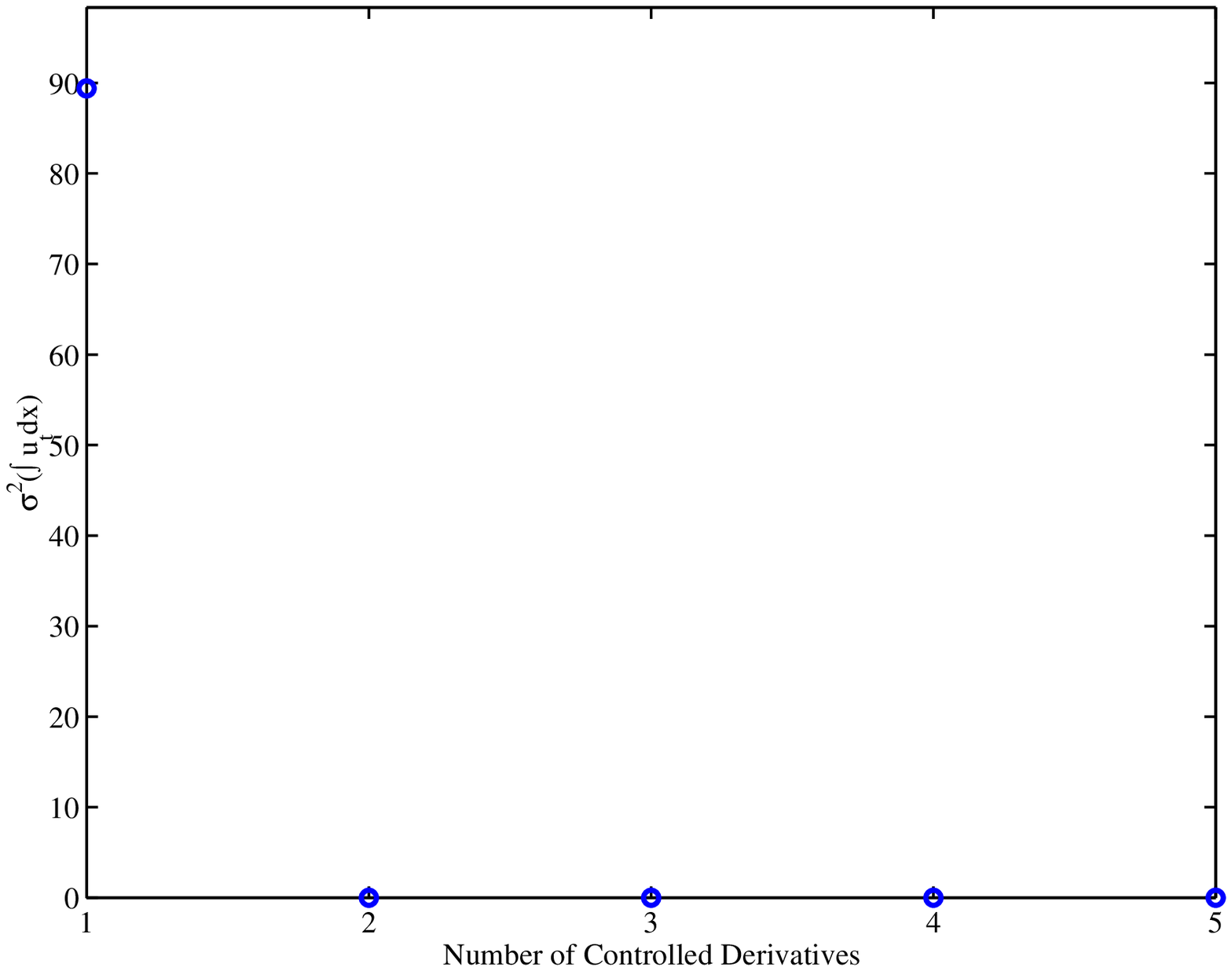, width=6.2cm,angle=0, clip=}
(a)
\epsfig{file=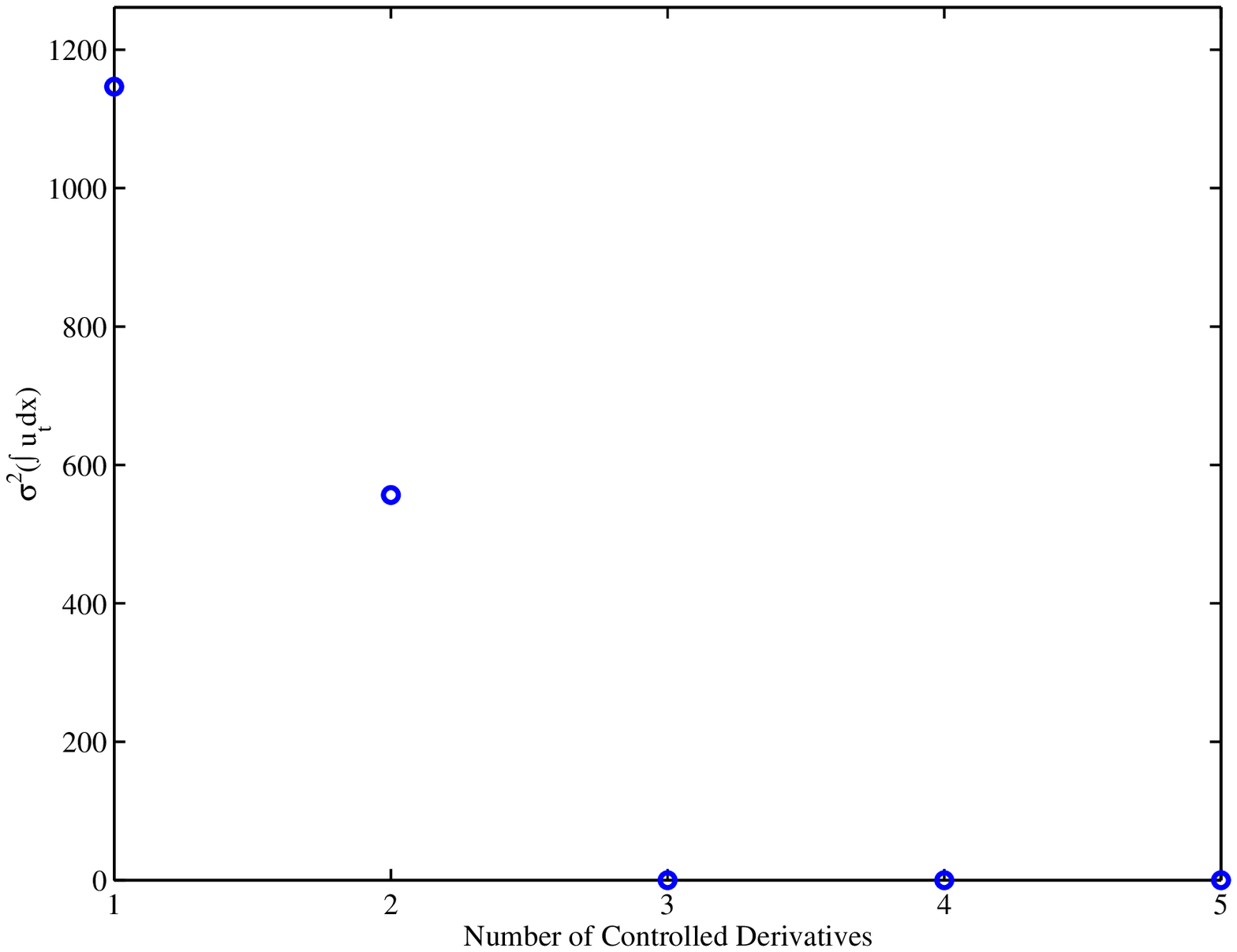, width=6.2cm,angle=0, clip=}
(b) }
\caption{(a) Conservation of the Burgers finite-difference PDE
time-stepper (\ref{BurgersFiniteDifferenceIntegrator}). (b)
Conservation of the KdV finite-difference PDE time-stepper
(\ref{KdVFiniteDifferenceIntegrator}).}
\label{PDEConservation}
\end{figure}

Fig. \ref{BurgersPDEConservationCtrl} plots families of initial
profiles thus constructed with progressively more controlled initial
derivatives.
Fig. \ref{PDEConservation}(a) and \ref{PDEConservation}(b) show the
results of applying the identification scheme to the Burgers
finite-difference PDE time-stepper
(\ref{BurgersFiniteDifferenceIntegrator}) and the KdV
finite-difference PDE time-stepper
(\ref{KdVFiniteDifferenceIntegrator}), respectively. $N^\prime$ is
identified to be $1$ for (\ref{BurgersFiniteDifferenceIntegrator}) and
$2$ for (\ref{KdVFiniteDifferenceIntegrator}).

Two comments are in order: first, we probe the consequences of
conservation (i.e. that boundary fluxes are the only cause of change
for the conserved quantity in a domain); second, we obtain (as a
side-product) the highest spatial derivative of the conserved quantity
$u$ in the constitutive equation for the flux.
It is important to note that if the procedure progressively returns
negative answers (e.g., if the sample variance is non-zero for a given
number $n$ of controlled derivatives), 
this does {\it not} imply that a conservation law does not
exist.
It only implies that a conservation law of the class encompassed by
our equation (\ref{LocalEvolutionEquation}) with spatial derivatives
up to the tested order $n$ does not exist.
In that sense, our procedure provides sufficient confirmation, but its
success is not necessary for a conservation law in a different class
to prevail.
Nevertheless, the class we consider is wide enough to encompass many
known examples and problems of interest to applications.

Note that $N^\prime$ is one order less than $N$ identified in section
\ref{sec:Identification} in both cases.
This is true  in 1-D because of 
Eq. (\ref{FluxExistence}).  %
So in 1-D, the baby-bathwater schemes give a definite answer to
whether $u$ is conserved {\em or not} in a finite $N-1$ steps, as long
as $N$ is identified first.

\section{Discussion}

In section \ref{sec:Conservation} we proposed methods to check whether
a coarse quantity (such as the mass corresponding to the coarse density
$g(u,u_x,u_{xx},..,u_x^{(N^{\prime\prime})})$) is conserved, without
knowledge of the coarse evolution equation.
The obvious question that arises is how do we know which $g$ to check?
The path that we suggest here, in the equation-free setting, is to
examine the {\it consequences} of conservation laws.
For example, consider the conservation of (linear) momentum.
An equivalent statement, through Noether's theorem \cite{Arnold89} is
the existence of translational invariance.
If we numerically establish the latter, then we can claim the former.
Let us then consider initial conditions to the available integrator
which are {\it shifts} of an original profile e.g., $u(x-x_0)$,
$u(x-(x_0+\epsilon))$, $u(x-x_0+2 \epsilon)$, etc.
Then if we time evolve the problem, using our microscopic
time-stepper, and the equation is translationally invariant, upon
reaching the integration reporting horizon, we can back-shift the
profile (by the original shift amount).
If all back-shifts provide an identical profile, we can conclude
translational invariance and hence linear momentum conservation.
An additional note of caution is that the examination of such
consequences is relevant when Noether's theorem applies, hence when
there is an underlying Lagrangian/Hamiltonian structure in the problem
(we discuss separately the issue of Hamiltonian nature below).
Notice, however the modulo the proviso of ``Hamiltonianity'', this
methodology can be used to establish additional dynamical invariants,
e.g. the invariance with respect to phase of the evolution of a field
can be related to norm invariance etc.

Establishing an underlying Hamiltonian structure in a sense proceeds
in a similar fashion through its correlation with invariance with
respect to time reversal.
The crudest way to examine this is by simply running the integrator with a
negative time-step (if that option is available).
A more refined way to check the same symmetry is by examining
computations of the spectrum (e.g., eigenvalues) of linearization of
the coarse PDE.
In particular, a straightforward consequence of the Hamiltonian nature
is that all linearization eigenvalues should come in quartets, namely
if $\lambda$ is an eigenvalue, then so are
$-\lambda,\lambda^{\star},-\lambda^{\star}$, where $^{\star}$ denotes
complex conjugation.
It is fortunate that time-stepper based numerical analysis techniques
for the numerical approximation of the leading spectrum of such a
linearization are well-developed for the case of {\it large scale}
continuum simulations (see for example
\cite{Christodoulou88,Goldhirsch87,Arnoldi51,Lanczos50,Tuckerman99,Bai00}).
If an eigenvalue $\lambda$ of the linearization is identified,
matrix-free eigen-computations with shift can be used to explore the
existence of the $-\lambda$ eigenvalue (in general, real eigenvalues
will come in pairs and complex conjugate eigenvalues in quartets).

While coarse time-reversibility can be answered by exploring the
spectrum of the linearization, it raises the interesting question of
how to integrate backward in time with the microscopic code.
Consider a molecular dynamics configuration with a certain set of
velocities at time zero, and the same molecular configuration with
``flipped" velocities.
A well known (and testable) consequence of microscopic
reversibility and the ``molecular chaos" ansatz is that, whether we
integrate the molecular dynamics equations forward or backward in time
starting from a randomly picked phase point, we will get ``the same"
forward in time evolution of the coarse macrosocpic observables.
It is interesting, however, that the coarse projective integration
techniques in an equation-free context {\it can} be used to attempt
integration of the {\it coarse} variables backward in time (under
appropriate conditions about the spectrum of the unavailable equation)
as follows:
Consider the lifting of a particular coarse initial condition to
consistent molecular realizations; flipping the molecular velocities
for these realizations will {\it not} affect the coarse procedure.
We then evolve microscopically the molecular configurations (whether
with the original or with flipped velocities) forward in time long
enough for the higher moments to heal, say for a time $\tau >
\tau_{mol}$.
We now estimate the time derivative $du/dt$ of the healed coarse
variables from the restriction of the ``tail end" of the molecular
trajectories, and then take a {\it large}, macroscopic Euler step {\it
backward in time} for the coarse variables.
We lift again, run molecular dynamics forward or backward
microscopically, estimate the coarse forward time derivative, and take
another coarse backward step.
This procedure can of course be done in a much more sophisticated way
as far as the coarse backward time step is concerned - algorithms like
Runge-Kutta or Adams method can be combined with the MD computations
to integrate density expectations {\it backward in time} for the
coarse equations on the ``slow manifold".
This ``see-saw" forward-backward coarse integration procedure can also
be used on stiff systems of ODEs and even dissipative PDEs under the
appropriate conditions to evolve trajectories backwards {\it on a slow
manifold}.
The numerical analysis of these algorithms in the continuum case is
an interesting subject in itself, and we are currently pursuing it
\cite{Gear03}.
It is interesting that the technique, in the molecular dynamics case,
can be used to coarsely integrate backward in time on a free energy
surface, and thus help molecular simulations escape from free energy
minima; we have already confirmed this in the case of Alanine
dipeptide folding in water at room temperature through molecular
dynamics simulations \cite{Hummer02}.

Finally, a more complicated question than checking the existence of
one (a specific, and hence related to a specific invariance, in
accordance with the above discussion) integral of the motion is the
one of integrability.
The latter necessitates infinite integrals of the motion, normally
established by means of identifying Lax pairs and using the inverse
scattering transformation machinery \cite{Ablowitz81}.
However, one can also use in this case consequences of integrability
to establish it.
For instance, in recent work \cite{PKevrekidis02} it was qualitatively
argued (and verified through numerical experiments in different
settings) that a feature particular to integrable Hamiltonian systems
is the presence of double continuous spectrum eigenvalues, when
linearizing around a (coarse PDE) solitary wave under periodic
boundary conditions.
These as well as other criteria (such as the existence of point
spectrum eigenvalues in the spectral gap \cite{PKevrekidis01}) can
also be (conversely) used to potentially rule out the existence of
integrable structure.
In short, the spectral properties of the coarse PDE linearization can
be used to establish or disprove not only the Hamiltonian (see above),
but also potentially the integrable nature of the flow.
While these are just initial thoughts towards attempting to decide
vital questions about the nature of the unavailable closed equation,
it is important to note that, what is computationally involved is a
time-stepper based identification of facts about the spectrum of the
linearization of an operator.
This ``computational technology" is quite mainstream in the case of
large scale continuum simulators, and 
can be straightfowardly
adapted to the case of coarse timesteppers in conjunction with the
lifting-restriction steps.
Variance reduction will clearly be the most significant step in the
wide applicability of these and similar-spirited approaches.

{\bf Acknowledgements} The authors would like to thank Frank Alexander
for stimulating discussions. This work was partially supported by
the National Science Foundation (IGK,PGK) AFOSR (CWG,IGK) and the
Clay Institute (PGK).

\appendix

\section{Coarse Density Lifting / Restriction Operators}
\label{sec:density}

For a coarse field ${\bf u}$, the lifting operator $\hat{\mu}$
generates a microstate ${\bf U}$: ${\bf U}=\hat{\mu}{\bf
u}$. Similarly, for a microstate ${\bf U}$, the restriction operator
$\hat{\cal M}$ returns a coarse field estimate $\tilde{\bf u}$:
$\tilde{\bf u}=\hat{\cal M}{\bf U}$. Both $\hat{\mu}$ and $\hat{\cal
M}$ can be one-to-one or one-to-many operators, but we demand that
$\hat{\cal M}\hat{\mu}\rightarrow \hat{I}$ asymptotically when the
wavelength of ${\bf u}$ is large enough compared to the microscopic
length scale \cite{Gear02a}. For 1-D coarse density field $u(x)$ under
$x\in[0,2\pi)$ PBC, we use the following $\hat{\mu}$, $\hat{\cal M}$
operators for the sake of definiteness in numerical experiments, even
though their construction is not unique.

The lifting operator $\hat{\mu}$, $u(x)\rightarrow \{x_i\}$:
\begin{enumerate}
\item Estimate $u_{\rm max}^{\rm approx}\approx u_{\rm
max}\equiv\max_{x\in[0,2\pi)} u(x)$. Pick $u_{\rm safe}$ that is
``safely'' greater than $u_{\rm max}$, for example $u_{\rm
safe}=1.1u_{\rm max}^{\rm approx}$.

\item Define $N^\prime\equiv \lceil 2\pi u_{\rm safe} Z\rceil$.
Create $N^\prime$ particles $\{x_i\}$ with each $x_i$ independently
drawn from uniform distribution on $[0,2\pi)$.

\item Go to each particle $i$, randomly decimate it with probability
$1-\frac{2u(x_i)}{u_{\rm max}^{\rm approx}+u_{\rm safe}}$. Count the
total number of surviving particles $N^{\prime\prime}$.

\item Compute quadrature,
\begin{equation}
Q\equiv Z\int_0^{2\pi} u(x)dx
\end{equation}
randomly round to $N=\lceil Q \rceil$ or $N=\lceil Q \rceil+1$ such
that $\langle N \rangle = Q$. Randomly pick $N^{\prime\prime}-N$
particles out of the $N^{\prime\prime}$ survivors and decimate them.
We now have a set of particles $\{x_i\}$, totally numbered either
$\lceil Q \rceil$ or $\lceil Q \rceil+1$.
\end{enumerate}

The restriction operator $\hat{\cal M}$, $\{x_i\}\rightarrow
\tilde{u}(x)$:
\begin{enumerate}
\item Define microscopic density function,
\begin{equation}
a(x)\equiv \frac{1}{Z} \sum_{i=1}^N \delta(x-x_i^-)
\end{equation}
and corresponding cumulant function,
\begin{equation}
c(x)\equiv \int_0^x a(x^\prime)dx^\prime.
\end{equation}
Clearly, at the the first, second, third particle positions $x_{n_1}$,
$x_{n_2}$, $x_{n_3}$, $c(x_{n_1})=1/Z$, $c(x_{n_2})=2/Z$,
$c(x_{n_3})=3/Z$, etc. And we have $c(0)=0$, $c(2\pi)=N/Z$.

\item Define a residual function $r(x)$,
\begin{equation}
r(x)\equiv c(x) - \frac{Nx}{2\pi Z}
\end{equation}
which is the difference between $c(x)$ and the cumulant of a
homogenized particle gas background. The idea is that
$r(0)=r(2\pi)=0$, so it is a periodic function and can be approximated
by,
\begin{equation}
r(x)\approx \tilde{r}(x) =
\sum_{i=1}^M a_n(\cos(nx)-1) + b_n \sin(nx).
\end{equation}
In fact, a sound strategy is to least-square fit $\tilde{r}(x)$ (its
$\{a_n\}$,$\{b_n\}$ coefficients) to $r(x)$ at $x=x_{n_i}$'s, where
$\{x_{n_i}\}$ is the sorted list of $\{x_i\}$. $r(x)$ can be easily
evaluated at $x_{n_i}$'s, noting the last sentence of step 1.

\item The coarse density estimate can be obtained by taking the
derivative of $\tilde{c}(x)\equiv Nx/2\pi Z + \tilde{r}(x)$,
\begin{equation}
\tilde{u}(x) = \frac{N}{2\pi Z} + 
\sum_{i=1}^M - na_n\sin(nx) + nb_n \cos(nx).
\end{equation}
\end{enumerate}

It is worth noting that although the constructed $\hat{\cal M}$
depends on $M$, it satisfies the particle number conservation {\em
exactly} because the finite harmonics all integrate to zero and only
the background contribution remains. In fact, $\langle \hat{\cal
M}\hat{\mu} \rangle$ also satisfies exact particle number conservation
to the original $u(x)$ under probabilistic average. Further, one can
show $\langle\hat{\cal M}\hat{\mu}\rangle= \hat{I}$ exactly for $u(x)$
in the first $M$ harmonics subspace.

\end{document}